\documentclass[pra, aps, twocolumn, preprintnumbers, superscriptaddress]{revtex4}
\usepackage{verbatim}
\usepackage{amsmath}
\usepackage{latexsym}
\usepackage{revsymb}
\usepackage{yfonts}

\usepackage{natbib}
\usepackage{amsfonts}
\usepackage{amsmath}
\usepackage{amssymb}
\usepackage{amsthm}
\usepackage{graphicx}
\usepackage{bm}
\usepackage{bbm}
\usepackage{multirow}

\newcommand{\be}{\begin{eqnarray} \begin{aligned}}
\newcommand{\ee}{\end{aligned} \end{eqnarray} }
\newcommand{\benn}{\begin{eqnarray*} \begin{aligned}}
\newcommand{\eenn}{\end{aligned} \end{eqnarray*} }

\newcommand{\bc}{\begin{center}}
\newcommand{\ec}{\end{center}}

\newcommand{\id}{\mathbb{I}}

\newcommand{\Tr}{\mathop{\mathrm{Tr}}\nolimits}

\newtheorem{theorem}{Theorem}[section]

\newtheorem{lemma}[theorem]{Lemma}

\newtheorem{corollary}[theorem]{Corollary}

\newcommand{\hil}{\mathcal{H}}

\renewcommand{\H}{\operatorname{H}} 
\newcommand{\hmin}{\ensuremath{\H_{\infty}}}
\newcommand{\hmine}[2]{\ensuremath{\hmin^{#1}\left(#2\right)}}
\newcommand{\hminee}[1]{\hmine{\varepsilon}{#1}}
\newcommand{\ball}{\mathcal{K}^\eps}

\newcommand{\set}[1]{\{#1\}}
\newcommand{\bop}{\mathcal{B}}
\newcommand{\pos}{\mathcal{P}}

\usepackage{amsfonts}

\def\id{\mathbb{I}}

\def\01{\{0,1\}}

\newcommand{\eps}{\varepsilon}
\newcommand{\ket}[1]{|#1\rangle}
\newcommand{\bra}[1]{\langle#1|}
\newcommand{\outp}[2]{|#1\rangle\langle#2|}
\newcommand{\proj}[1]{|#1\rangle\langle#1|}

\newcommand{\inp}[2]{\langle{#1}|{#2}\rangle} 

\newcommand{\rank}{\operatorname{rank}}

\newcommand{\mX}{\mathcal{X}}

\newcommand{\setA}{\mathcal{A}}
\newcommand{\setB}{\mathcal{B}}

\newcommand{\setS}{\mathcal{S}}
\newcommand{\setD}{\mathcal{D}}
\newcommand{\setG}{\mathcal{G}}

\newcommand{\setM}{\mathcal{M}}

\newcommand{\setT}{\mathcal{T}}
\newcommand{\setX}{\mathcal{X}}

\newcommand{\states}{\mathcal{S}}
\newcommand{\assign}{:=}
\newcommand{\setR}{R}

\begin{document}

\title{A lower bound on the dimension of a quantum system given measured data}

\author{Stephanie \surname{Wehner}}
\affiliation{Institute for Quantum Information, 
California Institute of Technology, 
1200 E California Blvd, Pasadena CA 91125, USA}
\author{Matthias \surname{Christandl}}
\affiliation{Arnold Sommerfeld Center for Theoretical Physics, 
Faculty of Physics,
Ludwig-Maximilians-University Munich, 
Theresienstr. 37, 80333 Munich, Germany}
\author{Andrew C. \surname{Doherty}}
\affiliation{School of Physical Sciences, University of Queensland, Queensland 4072, Australia}

\date{\today}
\begin{abstract}
We imagine an experiment on an unknown quantum mechanical system in
which the system is prepared in various ways and a range of
measurements are performed. For each measurement $M$ and
preparation $\rho$ the experimenter can determine, given enough time,
the probability of a given outcome $a$:  $p(a|M,\rho)$. 
How large does the Hilbert space of
the quantum system have to be in order to allow us to find density matrices and
measurement operators 
that will reproduce the given probability distribution?
In this note, we prove a simple lower
bound for the dimension of the Hilbert space. The main insight is to
relate this problem to the construction of quantum random access
codes, for which interesting bounds on the Hilbert space dimension already exist.
We discuss several applications of our result to hidden variable, or
ontological models, to Bell inequalities and to properties of the
smooth min-entropy.
\end{abstract}
\maketitle

Loosely speaking, the dimension of the Hilbert space describing a
quantum mechanical
system limits the complexity or usefulness of the correlations
that can be generated by experiments on the system. 
For example, it has been suggested that the primary resource for
quantum computation is Hilbert space
dimension~\cite{blume-kohout}. 
In practice though, when an experimentalist is faced with a real physical
system the dimension of the Hilbert space is often infinitely large in
principle. The dimension of the Hilbert space that we use to describe
the system of interest usually 
depends on the approximation used to describe the physics of the
system and may well depend on how well the experiment has in fact been
set up. For this reason it would be of interest to be able to use the
correlations observed in experiment to find strict lower bounds on the
dimension of Hilbert space. Thus one could conclude based on
experimental data that the Hilbert space
dimension of some system of interest was necessarily large and that
the system could not be effectively approximated by a smaller one.
In this paper we show that it is certainly possible to derive
very general lower bounds on Hilbert space dimension given experimental
data. 

The properties of quantum correlations have been best studied in the
setting of the Bell experiment.
Imagine two parties, Alice and Bob, who are given access to shared quantum states $\ket{\Psi}_{AB}$, but cannot
communicate. Each of them now performs a randomly chosen measurement on $\ket{\Psi}_{AB}$ and records their measurement
outcome. In order to obtain an accurate estimate for the correlation between their choice of measurement settings
and measurement outcomes, Alice and Bob now perform this experiment many times, using an identically prepared state $\ket{\Psi}_{AB}$
in each round. Quantum mechanics imposes strict limits on the strength of such non-local correlations, and it has been shown
that we can compute bounds on these correlations for any such
experiment~\cite{acin:bell,acin:bell2,qmp}. (These bounds generalize
the well known Tsirelson
inequalities~\cite{tsirel:original,tsirel:hadron} that apply to 
conventional Bell experiments that test the
Clauser-Horne-Shimony-Holt inequality.)
In particular, if we let $p(a,b|s,t)$ be the probability that Alice and Bob obtain measurement outcomes $a \in A$ and $b \in B$ 
when performing measurements indexed by $s \in S$ and $t \in T$, we can test using the methods of~\cite{acin:bell,acin:bell2,qmp} whether there exists a shared state $\ket{\Psi}$
and measurement operators $M_s^a$ and $M_t^b$ for Alice and Bob such that
$$
p(a,b|s,t) = \bra{\Psi}M_s^a \otimes M_t^b \ket{\Psi},
$$
for all $a$,$b$,$s$, and $t$. 
But how large does the dimension of the Hilbert space have to be such that we can find such a state and measurements?

Unfortunately, the methods of~\cite{acin:bell,acin:bell2,qmp} do not give us any bound on the dimension
in general. It is known that in the special case of two-party
correlations, where Alice
and Bob perform measurements using observables with eigenvalues $\pm 1$ (also known as XOR-games with $A = B = \01$), 
the dimension of the entangled state does not need to be larger than $d = 2^n$, where $n = \min(|S|,|T|)$~\cite{tsirel:original,tsirel:hadron}. Results are also known for a certain sets of 2-outcome inequalities~\cite{vertesi:efficiency,vertesi:dimension}. 
Very little is known otherwise. Even though one can construct an inequality with an infinite number of settings that requires an infinitely large entangled 
state~\cite{hypercube}, 
it is unknown whether there exist general experiments with a \emph{finite} number of measurement
settings for which an infinitely large entangled state is required to obtain the maximum possible quantum violation exactly.

In the context of bipartite Bell experiments, the question of determining the Hilbert space dimension from experimental data has
been addressed in the recent work of Brunner et
al.~\cite{brunner:dwitness}. 
Their aim was to reproduce the statistics of an
experiment performed by two separated observers on a single preparation
of a
bipartite quantum system. They introduce the concept of dimension
witness, which is a certain kind of generalization of Bell inequalities that
make it possible to distinguish the strength of correlations that can be
obtained in different dimensions.
This very nice approach makes it possible to find interesting lower bounds on
the dimension
of the system in use and has recently been extended by Bri{\"e}t et al.~\cite{jop:grothendieck} for XOR games. Our work finds rather different bounds on
Hilbert space dimension that are obtained by a very different
method. The bounds
apply to quantum mechanical systems with any number of parties (even one), and
apply also to the case where the experimental data refers to an arbitrary
number of preparations of the system.
Our bound is of
particular significance if the number of measurement
outcomes for each party is small.

The general problem we consider is this:
Suppose we are given a set of preparations
$\setS$ of a given quantum system and a set of measurements $\setM$,
each of which has outcomes $a \in \setA$~\footnote{W.l.o.g., we will take all measurements to have the same number of outcomes, as we may extend them otherwise}.  
We are given, perhaps as a
result of experiments, probabilities
$p(a|j,r)$
of obtaining outcome $a$ when performing the measurement $M_j \in \setM$
having prepared the system in state $r \in \setS$. However we do not
know either an explicit density matrix $\rho$ for the preparation $r$
or a measurement operator $M^a$ such that 
$$
p(a|m,r)={\rm Tr}(\rho M^a)=p(a|M,\rho)
$$
where
we use $M^a$ to denote the measurement operator corresponding to
outcome $a$ of measurement $j$, and will simply write $p(a|j,\rho) \assign p(a|j,r)$ from now
on. How large does the dimension of the Hilbert space supporting the
states $\rho$ have to be? 

This question was recently raised in~\cite{terry:ontArticle, terry:ontological}
which determined the number of (hidden) variables in an ontological
model necessary to reproduce the probabilities $p(a|j,r)$. 
In particular, it was shown that if each measurement has only two outcomes, then for a particular ontological model the
number of hidden variables must be greater than $N = \min(|\setS|, 2^{|\setM|})$.
Here, we prove a simple lower bound that shows that in the quantum setting the dimension of our space
scales as $2^{c \log N}$, where $c$ is a constant depending on the probabilities above.
Thus, if the number of states $|\setS|$ and the number of measurements $|\setM|$ 
is large, the dimension of the quantum state that we need cannot be
significantly smaller.

In the following, we first prove a simple lower bound for this general problem. We then examine how we can use this to lower bound
the dimension of the entangled state in a Bell experiment, and provide a simple example. In the appendix, we show that this example disproves that the smooth min-entropy
is additive and that we can perform exact min-entropy splitting as for independent states, which is of interest in the noisy-quantum-storage model~\cite{steph:diss,prl:noisy,noisy:robust}.

Throughout this note, we use $h(p) \assign -p \log p - (1-p) \log (1-p)$ to denote the binary entropy,
where all logarithms are taken to base 2. We furthermore use $\states(\hil)$ to denote the
set of all quantum states on the Hilbert space $\hil$, and write $H(\rho) \assign - \Tr(\rho \log \rho)$
for the \emph{von Neumann entropy} of a state $\rho \in \states(\hil)$. Note that if $\rho$ is classical, this reduces to 
the Shannon entropy, and that $\log(\dim(\hil)) \geq H(\rho) \geq 0$~\cite[Theorem 11.8]{nielsen&chuang:qc}, since
we may equivalently write $H(\rho) = - \sum_j \lambda_j \log \lambda_j$ where $\lambda_j$ is the $j$-th eigenvalue of $\rho$.
We will also need the concept of a \emph{cq-state} $\rho^{XQ} \in \states(\hil^X \otimes \hil^Q)$, a state that is part classical, part quantum, of the form
$$
\rho^{XQ} = \sum_{x \in \setX} P_X(x) \outp{x}{x} \otimes \rho^Q_x,
$$
where $P_X$ is a probability distribution over $\setX$ and $\inp{x}{x'} = 0$ for $x \neq x'$. 
Let $\rho^X = \Tr_Q(\rho^{XQ})$ and $\rho^Q = \Tr_X(\rho^{XQ})$
be the reduced states on systems $\hil^X$ and $\hil^Q$ respectively.
The \emph{conditional von Neumann entropy} is defined as $H(X|Q) \assign H(\rho^{XQ}) - H(\rho^{Q})$. 
We will also use the shorthands $H(X) \assign H(\rho^X)$ and $H(XQ) \assign H(\rho^{XQ})$
and $[\ell] \assign \{1,\ldots,\ell\}$.

\section{Lower Bound}
We first state the intuition behind our simple lower bound, based
on quantum random access codes. A quantum $(m,q,p)$-random access code is an encoding
of an $m$-bit string $x$ into a $q$-qubit state $\rho_x$ such that for any 
$i \in [m]$ we can retrieve the bit $x_i$ from $\rho_x$ with probability $p$.
Note that we are only interested in retrieving a single bit of the original string $x$
from $\rho_x$. In general, it is unlikely that we will be able to retrieve more than a single
bit. For such encodings it is not hard to prove a lower bound on the number of 
qubits $q$~\cite{nayak:rac} if the distribution over the strings is uniform and the 
probability of decoding each bit is the same.

Now note that our problem has a very similar flavor.
Suppose we were given states $\rho_1,\ldots,\rho_\ell$
and measurements $M_1,\ldots,M_m$ that give us the desired probabilities
$p(a|M_j,\rho_x)$. For simplicity, assume for now that $\ell = 2^m$ and $a \in \01$. 
Then the states $\rho_1,\ldots,\rho_\ell$ form a generalized quantum random access code, 
where each state represents an encoding of an $m$-bit string $x$ and we think of 
$M_j$ as the measurement that we can apply to extract bit $x_j$ with probability 
$p(a|M_j,\rho_x)$. Once we realize this viewpoint it is indeed very intuitive that we should be able to apply 
techniques similar to the ones used for quantum random access codes also in the present setting. 

\subsection{Tools}
We first state a general lemma from which our bound later follows by constructing an appropriate mapping that associates a string $x$
with a state $\rho_x$.
Our proof is a straightforward extension of the techniques employed
for the random access code lower bound~\cite{nayak:old,nayak:rac,kerenidis&wolf:qldcj} to more generalized distributions and alphabets:

\begin{lemma}\label{protoBound}
Let $\setX = \setA^{\times m}$ denote the set of strings of length $m$, let $P_X$ be a probability distribution over $\setX$
and let $X=X_1,\ldots,X_m$ denote a random variable chosen from $\setX$ according to the distribution $P_X$.
Let $\hil^Q$ be a Hilbert space supporting an 
an ensemble of states $\{ P_X(x), \rho_x \mid \rho_x \in \states(\hil^Q), x \in \setX\}$
and POVMs $E_j=(E_j^{z_j}), \ \ j \in [m]$ with outcomes
$z_j \in \setA$.
Let $Z_j$ be the random variable corresponding to the decoding of $X_j$ by performing the measurement $E_j$ on $\hil^Q$
where we use 
$P_{Z_j|X_j}(z_j|x_j) \assign \Tr(E_j^{z_j}\rho_x)$ 
to denote the conditional probability distribution of a random variable $Z_j$ over $\setA$. 
Then
$$
\dim(\hil^Q) \geq 2^{H(X) - \sum_j H\left(X_j|Z_j\right)}.
$$
\end{lemma}
\begin{proof}
Consider a cq-state $\rho^{XQ} \in \states(\hil^X \otimes \hil^Q)$ of the form
$$
\rho^{XQ} = \sum_x P_X(x) \outp{x}{x} \otimes \rho_x^Q,
$$
where $\rho_x^Q \assign \rho_x$.
We have
\begin{eqnarray*}
\log(\dim(\hil^Q)) \geq H(Q) &\geq& H(Q) - \sum_x P_X(x) H(\rho^Q_x)\\
        &=& H(X) + H(Q) - H(XQ)\\
        &=& H(X) - H(X|Q)\\
        &\geq&H(X) - \sum_{j=1}^m H(X_j|Q),
\end{eqnarray*}
where the first inequality follows from $H(Q) \leq \log(\dim(\hil^Q))$~\cite[Theorem 11.8.2]{nielsen&chuang:qc},
the second from the fact that for all $\rho^Q_x$ we have $H(\rho^Q_x) \geq 0$~\cite[Theorem 11.8.1]{nielsen&chuang:qc},
the third equality from $H(XQ) = H(X) + \sum_x P_X(x)
H(\rho^Q_x)$~\cite[Theorem 11.8.5]{nielsen&chuang:qc},
the fourth from the definition of the conditional von Neumann entropy, and the last inequality
from its strong subadditivity $H(X_1\ldots X_m|Q) = H(X|Q) \leq \sum_{j=1}^m H(X_j|Q)$~\cite[Theorem 11.16]{nielsen&chuang:qc}, where $X_j$ is the random variable corresponding to the $j$-th entry of $X$.

Finally, note that we can express the effects of a measurement $M$ on $\hil^Q$ by performing a unitary operation $U$
on $\rho^{XQ} \otimes \outp{z_0}{z_0} \in \states(\hil^X \otimes \hil^Q \otimes \hil^{Z_j})$ with $\outp{z_0}{z_0}$ an
initial pure state of $\hil^{Z_j}$, followed by tracing out the
ancilla $\hil^{Z_j}$ holding the measurement outcome~\cite{nielsen&chuang:qc}.
We then have $H(X_j|Q) = H(X_j|QZ_j)$ since $U$ is unitary, and $H(X_j|QZ_j) \leq H(X_j|Z_j)$ since conditioning reduces entropy~\cite[Theorem 11.15.1]{nielsen&chuang:qc},
from which the claim follows.
\end{proof}

This means that if we want to encode a string of $n$ dits~\footnote{A dit is a unit of information analogous to a bit that can take $|\setA| = d$ values.}
into a number
of qubits and attempt to recover the $j$-th dit with the $j$-th measurement,
then we need at least $H(X)-\sum_j H(X_j|Z_j)$ qubits. $H(X_j|Z_j)$
quantifies the uncertainty about the $j$-th bit given the outcome of the
$j$-th measurement. For instance, if the $n$ dits are drawn uniformly and
independently ($H(X) =n \log d$ where $d = |\setA|$) and we wish to recover them perfectly,
($H(X_j|Z_j)=0$ for all $j$) then we need $n \log d $ qubits to do so.
In the other extreme, where $Z_j$ holds no information about $X_j$ i.e. our recovery probability is no better than guessing,
we have $H(X_j|Z_j) = H(X_j) = \log d$ for all $j$ meaning that we need no qubits at
all for the encoding.

\begin{corollary}\label{simpleBound}
For the definitions as given in Lemma~\ref{protoBound}, it furthermore holds that
$$
\dim(\hil^Q) \geq 2^{H(X) - \sum_j \left(h(p_j) + (1-p_j)\log (|\setA|-1)\right)},
$$
where $p_j = \sum_{x \in \setX} P_{X}(x) \Tr(E_j^{x_j} \rho_x)$
is the average recovery probability of the $j$-th entry of $X$ when measuring $E_j$ on $\hil^Q$.
\end{corollary}
\begin{proof}
The statement follows immediately from Lemma~\ref{protoBound} and Fano's inequality giving
$H(X_j|Z_j) \leq h(p_j) + (1-p_j)\log(|\setA| - 1)$,
where $p_j$ is the average probability of correctly decoding the $j$-th bit of $X$ given access to
$Z_j$.
\end{proof}

Note that the bound further simplifies to
$$
\dim(\hil) \geq 2^{m(1-h(p))} 
$$
in the case where $X_j$ is binary and $P_X$ is the uniform distribution for which
$H(X_j)=1$ for all $j$ and the recovery probability for each bit is lower bounded by $p\geq \frac{1}{2}$. This last bound was first noted in the context of random access codes.
Lemma~\ref{protoBound} does in general give a better bound than Corollary~\ref{simpleBound}, although it may be harder to apply since it requires
more information about the distributions and will be less convenient for us when considering non-local games where we may have limited information.
Fano's inequality is tight for a distribution where the most likely outcome has probability $p_j$ and all
others have probability $p_j/(|\setA|-1)$ and in this case Corollary~\ref{simpleBound} gives exactly the same bound as Lemma~\ref{protoBound}.

\subsection{Dimension bound}

We are now ready to use these tools to prove a lower bound for our problem.
Intuitively, we let the states $\rho_1,\ldots,\rho_\ell$ corresponding to the preparations $r_1,\ldots,r_\ell$ represent
encodings of $m$-element strings $x$ chosen according to a probability distribution
$P_X$ from $\setX \assign \setA^{\times m}$. If $\ell < |\setA|^m$, then this just means that some strings have
0 probability of occurring. If $\ell > |\setA|^m$, then there are more elements in our 
string than we wish to extract in which case our lower bound will not be any stronger
than could be obtained by letting $\ell = |\setA|^m$. 
There is some freedom in applying the above bound to our setting, since we are
in general free to associate strings with states in any way we like, pick any of our available measurements $M_j$ to decode $x_j$ and finally
we may also choose any prior distribution $P_X(x)$, since our bound should hold for any such prior.
First, we associate strings
$x \in \setX$ with states $\rho \in \setS$ by constructing a map $g_{T,R}$ as follows:
Recall that without loss of generality, we
may order the states in lexicographic order $\rho_1,\ldots,\rho_\ell$. Let $T \subseteq [\ell]$ such that 
$|T| = \min(\ell,|\setX|)$, 
let $\setR \subseteq \setX$ such that $|\setR| = |T|$, and consider the set of one-to-one maps
$$
\setG_{T,\setR} \assign \{g_{T,\setR}: \setR \rightarrow T|
\forall x\neq x' \in \setX, g_{T,\setR}(x) \neq g_{T,\setR}(x')\}.
$$
That is, any map associates a unique state $\rho_{g_{T,\setR}(x)}$ with each string $x$.
Second, we now construct maps $e: [\ell] \rightarrow [\ell]$ and $c: \setA \times [\ell] \rightarrow \setA$,
that specify which measurement $E_j = M_{e(j)}$ we will use to extract a particular entry $x_j$ of
$x=x_1,\ldots,x_m$ from $\rho_{g_{T,\setR}(x)}$, for a potential relabeling of the outcomes as $M_{e(j)}^{c(a,j)} = E^a_j$ given by the map $c$.
Let $\setD = \{(e,c)\}$ denote the set of all such collections of maps.
Finally, we may choose $P_X$ to be any distribution over $\setX$, where we will assign probability $P_X(x) = 0$ to any $x \notin \setR$. 
Note that this means $P_X$ is effectively a distribution over $\setR$. 
If we take $P_X = P_{X_1} \times \ldots \times P_{X_m}$ to be a product distribution over all strings, then $e(j) = j$ is simply the identity, i.e., 
we will use measurement $M_j$ to decode the $j$-th element of the string.

We first of all show that Lemma~\ref{protoBound} gives us a lower bound on the dimension of the quantum system for \emph{any} distribution 
$P_X$, $T \subseteq [\ell]$, $\setR \subseteq \setX$ and mapping $g_{T,\setR}$, $c$, and $e$. This will be important in Section~\ref{sec:nonlocal}, where such mappings
are fixed when considering a particular non-local game. We state both consequences of Lemma~\ref{protoBound} and Corollary~\ref{simpleBound} explicitely:

\begin{corollary}\label{cor:protoBound}
Let $\setS = \{\rho_1,\ldots,\rho_\ell \mid \rho_j \in \states(\hil^Q)\}$ be a set of states and let $\setM = \{M_1,\ldots,M_m \mid M_j = (M_j^a) \in \bop(\hil^Q), a \in \setA\}$
be a set of POVMs satisfying $p(a|j,r) = \Tr(M_j^a \rho_r)$ for some given set of probabilities $\{p(a|j,r)\mid a \in \setA, j \in [m], r \in [\ell]\}$. 
Then for any $T \subseteq [\ell]$, $\setR \subseteq \setX$
with $|\setR| = |T| = \min(\ell,|\setX|)$, and
$g_{T,\setR} \in \setG_{T,\setR}$, $(e,c) \in \setD$, and any distribution $P_X$ over $\setR \subseteq \setX = \setA^{\times m}$ giving 
ensemble $\{P_X(x),\rho_{g_{T,\setR}(x)}\}$ we must have
$$
\dim(\hil^Q) \geq 2^{H(X) - \sum_j H\left(X_j|Z_j\right)},
$$
where $P_{Z_j|X_j}(z_j|x_j) \assign \Tr(M_{e(j)}^{c(x_j,j)}\rho_{g_{T,\setR}(x)})$.
Furthermore,
$$
\dim(\hil^Q) \geq 2^{H(X) - \sum_j \left(h(p_j) + (1-p_j)\log (|\setA|-1)\right)},
$$
where $p_j 
= \sum_{x \in \setX} P_{X}(x) \Tr(M_{e(j)}^{c(x_j,j)} \rho_{g_{T,\setR}(x)})$
is the average recovery probability of the $j$-th entry of $X$ when measuring $M_{e(j)}$ on $\hil^Q$.
\end{corollary}

We are now ready to state our main result as an immediate consequence of Corollary~\ref{cor:protoBound}.

\begin{theorem}\label{thm:dimBound}
Let $\setS = \{\rho_1,\ldots,\rho_\ell \mid \rho_j \in \states(\hil^Q)\}$ be a set of states and let $\setM = \{M_1,\ldots,M_m \mid M_j = (M_j^a) \in \bop(\hil^Q), a \in \setA\}$
be a set of POVMs satisfying $p(a|j,r) = \Tr(M_j^a \rho_r)$ for some given set of probabilities $\{p(a|j,r)\mid a \in \setA, j \in [m], r \in [\ell]\}$. 
Then
$$
\dim(\hil^Q) \geq 2^{C},
$$
with
$$
C \assign \max_{T,\setR,g_{T,\setR}(e,c),P_X} H(X)  - \sum_j H(X_j|Z_j)
$$
where the maximization is
taken over all 
subsets $T \subseteq [\ell]$, $\setR \subseteq \setX$
with $|\setR| = |T| = \min(\ell,|\setX|)$, and
probability distributions $P_{X}$ over $\setR$,
and mappings $g_{T,\setR} \in G_{T,\setR}$, $(e,c) \in \setD$,
with $P_{Z_j|X_j}(z_j|x_j) \assign \Tr(M_{e(j)}^{c(x_j,j)}\rho_{g_{T,\setR}(x)})$.
\end{theorem}

Note that if we fix $T$, $\setR$, $g_{T,\setR}$ and 
$(e,c)$ and furthermore restrict the maximization to product distributions $P_X = P_{X_1} \times \ldots \times P_{X_m}$
we have from Lemma~\ref{protoBound} combined with Corollary~\ref{cor:protoBound} that
$$
\dim(\hil^Q) \geq 2^{\sum_j C_j},
$$
where $C_j = \max_{P_{X_j}} I(X_j;Z_j)$ is the Shannon channel capacity, and $I(X_j;Z_j) = H(X_j) - H(X_j|Z_j)$ is the 
mutual information. Unfortunately, we do not know how hard it is to evaluate the quantity $C$ in general when maximizing over
all parameters. However, since for fixed $T,\setR,g_{T,\setR}, (e,c)$ it is equivalent to computing the Shannon channel capacity it may
not be an easy task for arbitrary distributions $p(a|j,r)$.

Let's look at a very simple example taken from~\cite{terry:ontArticle,terry:ontological}, that illustrates our bound.
The entries of the following table correspond to the probabilities $p(a|M_j,\rho_x)$, for
the two possible states labeled using strings '00' and '11'.

\begin{center}
\begin{tabular}{|c|c|c|c|}
\hline
$M$ & $a$ & $\rho_{00}$ & $\rho_{11}$\\
\hline
\multirow{2}{*}{$M_0$} & 0 & 1 & 0\\
& 1 & 0 & 1\\
\hline
\multirow{2}{*}{$M_1$} & 0 & 1 & 1/2\\
& 1 & 0 & 1/2\\
\hline
\end{tabular}
\end{center}

Note that in this example $\setX = \01^2$ consists of the possible strings of two bits, but
only $'00'$ and $'11'$ occur with non-zero probability. 
For simplicity, suppose we are given these states with probability $1/2$ each, and hence we have
$H(X) = 1$. Note that we can distinguish the two states perfectly
using the first measurement, and hence both encoded bits can be
recovered perfectly $p_1 = p_2 = 1$. By reference to Corollary I.2 we
see that at least a two-dimensional system is required to recover
these statistics. If we can only perform projective measurements,
then~\cite{terry:ontArticle,terry:ontological}
says that we need more than one qubit. Note however that this is
not the case for generalized measurements. To perform the second measurement $M_2$ we can perform the first
measurement $M_1 = \{M_1^a\}$, and output $0$ for outcome $0$ but for outcome $1$ we flip a coin that gives us $0$ and $1$
with probability $1/2$ each. This corresponds to letting $M_2^0 = M_1^0 + M_1^1/2$ and $M_2^1 = M_1^1/2$.
Hence, our bound is tight for this trivial example. Below, we provide a second
example that is inspired by the CHSH inequality.

Our analysis shows that it is indeed possible to obtain bounds on the dimension in the quantum setting, partially answering
an open question from~\cite{terry:ontArticle,terry:ontological} which asked to find such bounds for \emph{projective} 
measurements. In particular, note that if we choose a uniform prior over $N = \min(|\setS|,2^{\setM})$ possible states,
and consider only two outcome measurements we have by Corollary I.2 that the dimension of the system must obey 
$\log d \geq \sum_{j=1}^{\log N} (1 - h(p_j)) \geq \log N c$ with $c = \min_j (1 - h(p_j))$. This means that
in the case where $p_j$ is not arbitrarily close to $1/2$, and $N$ itself is very large, the dimension required
is not significantly different from the one required by the ontological model~\cite{terry:ontArticle,terry:ontological}.
It is worth considering the dependence on $p_j$ which seems to be absent from this particular ontological model. If we merely want
to represent the data classically in a way such that we can extract an arbitrary bit $x_j$ alone with probability $p_j=p$ and the prior distribution over
the strings is uniform, it is known that there do exist
classical random access codes for which the dimension obeys $\log d = \log N(1 - h(p)) + O(\log \log N)$~\cite{nayak:old}. 
Intuitively, the description of the ontological model includes much more information and hence has a larger size.
When examining information processing within such an ontological model, it may however be worth considering whether
it has a better representation for a particular task at hand.

\section{Non-local games}\label{sec:nonlocal}

We now show how our approach also leads to a lower bound on the dimension of the entangled state
that two or more parties need to share in \emph{any} Bell experiment, where we consider a bound for
the CHSH inequality as a small example. In this case we can immediately compute the lower bound since
all parameters $P_X$, $T$, $\setR$, $g_{T,\setR}$, and $(e,c)$ are fixed.
For the present purposes, it is convenient to view Bell experiments as a game between two, or more, distant players, who cooperate
against a special party. We call this special party the \emph{verifier}.
In a two player game with players Alice and Bob, the verifier picks two questions $s \in \setS$ and 
$t \in \setT$
and sends them to Alice and Bob respectively. Alice and Bob then return answers 
$a \in \setA$ and $b \in \setB$
to the verifier, who then decides according to a fixed set of public rules
whether Alice and Bob win by giving answers $a$ and $b$ to questions $s$ and $t$.
To win the game, Alice and Bob may agree on any strategy beforehand,
but can no longer communicate once the game starts. Classically, such
a strategy consists of shared randomness. In the quantum setting,
they may choose any entangled state as part of their strategy and agree on any measurements to be
performed on this state. Without loss of generality we can thus think of the questions as measurement 
settings and the answers as measurement outcomes. 

More formally, the game is characterized by finite sets $\setS, \setT, \setA, \setB$, 
a distribution $\pi: \setS \times \setT \rightarrow [0,1]$ according to which the verifier chooses his 
questions, and a predicate $V: \setA \times \setB \times \setS \times \setT \rightarrow \01$,
where $V(a,b|s,t) = 1$ if and only if $a$ and $b$ are winning answers given questions $s$ and $t$.
Let $\pi_A$ and $\pi_B$ be the marginal probability distributions over $\setS$ and $\setT$ respectively.
For simplicity, we also assume that we are dealing with a \emph{unique} game, where $V$ is defined
in such a way that for each $b,s,t$ there exists exactly one winning answer $a$ for Alice. 
Our argument for the general case is analogous, and can be obtained by combining the correct answers into one, which effectively corresponds
to performing a measurement with less outcomes. However, our proof just becomes much harder to read. For simplicity in our explanations, we will also assume
that the possible answers are the same for each possible measurement setting.

Let $\Pr[a|s]$ and $\Pr[b|t]$ be the probabilities that Alice and Bob return answers $a$ and $b$
given questions $s$ and $t$ respectively. Note that the no-signaling condition must hold and hence
we may without loss of generality assume these probabilities to be independent of the other parties measurement
setting.
We now show how to use our approach from above to lower bound the dimension of the entangled
state that Alice and Bob need to implement such a strategy. We are not concerned
with the question whether there actually exists a strategy for Alice and Bob to obtain said
distribution. This can be verified using the techniques of~\cite{qmp,acin:bell,acin:bell2}.

The simple trick is to realize that when Bob performs
a measurement on his part of the state, he prepares a certain state on Alice's end. 
Let $\chi_b^t$ denote the state that is prepared for Alice if Bob has measurement setting $t \in \setT$ and obtains outcome $b \in \setB$.  
The probability that Alice holds the state $\chi_b^t$ is given by 
$$ 
P_X(t,b) \assign \Pr[b|t]\pi_B(t),
$$
where we combine $t,b$ to index a string $x \in \setA^{|\setS|}$
as follows:
Note that since we are dealing with unique games, we can define a function $f: \setB \times \setS \times \setT \rightarrow \setA$ such that $f(b,s,t) = a$ for $V(a,b|s,t) = 1$. 
We can label Alice's measurements with numbers from one up to $|\setS|$ and hence
without loss of generality we will take $\setS = [|\setS|]$ to represent the set
of possible measurements for Alice. 
We define the string $x \in \setA^{|\setS|}$ as
\begin{equation}\label{eq:stringDef}
x \assign f(b,1,t),\ldots,f(b,|\setS|,t).
\end{equation}
and let 
$$
\rho_x \assign \chi_b^t. 
$$
Since $x$ is a function of $b$ and $t$, we have
$$
P_X(x) \assign P_X(t,b).
$$
If Alice chooses measurement setting $s$ she will
try and give the correct answer $a$. Note that effectively she tries to retrieve
the entry $x_s = f(b,s,t)$ from $\rho_x$, completing the analogy to quantum random access codes.

To apply Lemma~\ref{simpleBound},
let $p(a|M_s,\chi_b^t)$ be the probability that
Alice outputs $a$ for measurement setting $s$ and prepared state $\chi_b^t$.
\begin{corollary}\label{cor:nlGame}
In any non-local game where Alice obtains the correct outcome $a \in \setA_s$
for measurement setting $s \in \setS$ with probability $p_s$, the dimension of her
Hilbert space $\hil^A$ obeys
$$
\dim(\hil^A) \geq 2^{H(X) - \sum_s \left(h(p_s) + (1-p_s)\log (|\setA_s|-1)\right)}.
$$
where $X$ is the random variable corresponding to the choice of string as defined in Eq.~(\ref{eq:stringDef}).
\end{corollary}
Evidently, an analogous statement can be made for Bob. If we are considering more than two players,
it is straightforward to extend our argument to bound the Hilbert space dimension of each 
individual player by grouping the remaining players together as one.

Let's look at a small example
which illustrates the proof. 
Consider the CHSH game. Here, $\setA = \setB = \setS = \setT = \01$ and
$\pi$ is the uniform distribution.
Alice's goal is to obtain an outcome $a$ such that $s \cdot t = a + b \mod 2$. 
Letting $x = g(b,t) = f(b,0,t),f(b,1,t)$ we obtain an encoding of a two bit string $x \in \01^2$ as
$g(0,0) = 0,0$, $g(1,0) = 1,1$, $g(0,1) = 1,0$ and $g(1,1) = 0,1$. 
How many qubits does Alice need to use if she always wants to give the correct answer 
with probability $\gamma = 1/2 + 1/(2\sqrt{2})$? 
With analogy to the table of our previous example, we have probabilities $p(a|M_s,\rho_x)$ given by

\begin{center}
\begin{tabular}{|c|c|c|c|c|c|}
\hline
$M$ & $a$ & $\rho_{00}$ & $\rho_{01}$ & $\rho_{10}$ & $\rho_{11}$ \\
\hline
\multirow{2}{*}{$M_0$} & 0 & $\gamma$ & $\gamma$ & $(1-\gamma)$ & $(1-\gamma)$\\
& 1 & $(1-\gamma)$ & $(1-\gamma)$ & $\gamma$ & $\gamma$ \\
\hline
\multirow{2}{*}{$M_1$} & 0 & $\gamma$ & $(1-\gamma)$ & $\gamma$ & $(1-\gamma)$\\
& 1 & $(1-\gamma)$ & $\gamma$ & $(1-\gamma)$ & $\gamma$\\
\hline
\end{tabular}
\end{center}

We have for all $t,b \in \01$ $\Pr[b|t] = 1/2$ and hence
$P_X(x) = P_X(t,b) = 1/4$.
Since everything is uniform we immediately obtain from Corollary~\ref{cor:nlGame} that
$\log(\dim(\hil^A)) \geq (1 - H(p))2 \approx 0.8$. Hence
Alice needs at least one qubit to no great surprise.
We do not need to know a specific
strategy, however, for the well-known CHSH state and measurements we would have an encoding of 
$\rho_{00} = \outp{0}{0}$, $\rho_{01} = \outp{-}{-}$, 
$\rho_{10} = \outp{+}{+}$, and $\rho_{11} = \outp{1}{1}$ 
which actually coincides with the best known 
quantum random access code for a 2 bit string.

Bounds for other games for which we are given a distribution over the measurement outcomes can
be shown in an analogous way. In general, if we are given the full distribution over all settings
and outcomes we can apply the first part of Corollary~\ref{cor:protoBound} to obtain a slightly better bound,
depending on the distribution.  

\section{Min-entropy}

Our task of lower bounding the dimension of the Hilbert space can be
used to give a partial answer to an open problem in the
analysis of cryptographic protocols in the
bounded-quantum-storage~\cite{serge:bounded,serge:new}, and
noisy-quantum-storage
model~\cite{steph:diss,prl:noisy,noisy:robust}. In particular, the
example discussed in the previous section can be modified to give a
simple counterexample that shows that an additivity property of the
smooth min-entropy that has been shown to hold for independent 
quantum states~\cite{noisy:robust} is not true in general. 
Note that a modified version may still hold with additional loss in
the parameters.
The same counterexample can also be used to show that exact min-entropy splitting 
with respect to quantum knowledge as it holds for independent states~\cite{noisy:robust}
is not possible in general without imposing further assumptions.
We defer the details of this construction to the appendix.

\section{Conclusion}

We have given a simple lower bound that places a fundamental limit on how large the dimension of the state has to 
be to implement certain measurement strategies. Our result shows that in the limit of a large number of measurement
settings and states, the dimension of this state cannot generally be significantly smaller than the amount of classical
information (e.g. in the form of (hidden) variables in an ontological model~\cite{terry:ontArticle,terry:ontological}) necessary
to produce the desired statistics. 

Our approach also gives a weak bound on the dimension of the entangled state needed to implement non-local 
strategies for any multi-player non-local game. Note, however, that our bound will be quite weak if the probability
of outputting the correct outcome is close to $1/2$, or the number of measurement outcomes is large.
Furthermore, note that our bound also works for the case where the choice of Alice's measurement
settings is uniform which may not be the case for a particular game, leaving the possibility of better bounds.
Yet, our approach is a first direction
to find bounds for general games. It is an interesting
question whether the the present idea of viewing the game as an encoding procedure leads to new upper bounds as well.

\acknowledgments

We are indebted to the referee for helpful comments to improve the presentation of the paper.
ACD is supported by the Australian Research Council.
SW is supported by NSF grant number PHY-04056720. SW thanks the University of Queensland for the generous travel support 
to attend QIP '07, and Oscar Dahlsten and Renato Renner for the invitation to the very nice workshop at ETH Zurich in may 2008.

\appendix

\section{Min-entropy}

In this appendix, we describe the counterexample mentioned in the
text showing that the additivity property that was proved recently for
the smooth min-entropy 
of independent quantum states does not hold in general. 
The same example can also be used to show that min-entropy splitting with respect to 
quantum knowledge as it was shown for such states does not hold in general, without imposing additional constraints.

\subsection{Definitions}

To state the additivity lemma, we will need the following quantities introduced by
Renner~\cite{renato:diss}, reproduced here for convenience: Let $\rho_{AB} \in \states(\hil^A \otimes
\hil^B)$ and let $\sigma_B \in \states(\hil^B)$.  Then the
\emph{min-entropy of $\rho_{AB}$ relative to $\sigma_{B}$} is given by
$$
\hmin(\rho_{AB}|\sigma_B) \assign - \log \lambda \, ,
$$
where $\lambda$ is the smallest real number such that $\lambda \id_A
\otimes \sigma_B \geq \rho_{AB}$. We need a related quantity, where in addition
we optimize over states $\sigma_B$ defined as
$$
\hmin(\rho_{AB}|B) \assign \sup_{\sigma_B \in \states(\hil^B)}
\hmin(\rho_{AB}|\sigma_B) \, .
$$
For a cq-state $\rho_{XE}$, we also use the shorthand
$$
\hmin(X|E) \assign
\sup_{\sigma_E \in \states(\hil^E)} \H_\infty(\rho_{XE}|\sigma_E)
$$
for the \emph{conditional min-entropy of $X$ given $E$}. It is
difficult to get an intuitive understanding from this formal
definition of conditional min-entropy, but one can show using
semi-definite programming duality~\cite{KRS08} that
\begin{equation} \label{eq:duality}
\hmin(X|E) = - \log P_g(X|E) \, ,
\end{equation}
where $P_g(X|E)$ is defined as the maximum success probability of
guessing $X$ by measuring the $E$-register of
$\rho_{XE}$.
Formally, for any (not necessarily
normalized) cq-state $\rho_{XE}$, the guessing probability is defined as
\[ P_g(X|E) \assign \sup_{\set{M_x}} \sum_x P_X(x) \Tr(M_x \rho_E^x) \, ,
\]
where the supremum ranges over all positive-operator valued
measurements (POVMs) with measurement elements $\set{M_x}_{x \in
  \mX}$, i.e.~$M_x \geq 0$ and $\sum_x M_x = \id_E$.  If all side-information is
classical,
we recover the fact that the classical
min-entropy is the negative logarithm of the maximum probability.

We will also refer to smooth versions of these
quantities. Intuitively, we no longer consider the min-entropy of a
fixed state $\rho_{AB}$, but allow us to move to some
$\hat{\rho}_{AB}$ which is close to $\rho_{AB}$, but may have considerably
larger min-entropy. 
These smooth quantities are often needed since they have some nicer properties
than the conventional min-entropy.
For $\eps \geq 0$, the \emph{$\eps$-smooth
  min-entropy of $\rho_{AB}$ relative to $\sigma_{B}$} is given by
$$
\hminee{\rho_{AB}|\sigma_B} \assign \sup_{\hat{\rho}_{AB} \in
  \ball(\rho_{AB})} \hmin(\hat{\rho}_{AB}|\sigma_{B}) \, ,
$$
where $\ball(\rho_{AB}) \assign \{\hat{\rho}_{AB} \in \pos(\hil^A \otimes \hil^B)\mid \|\rho_{AB} - \hat{\rho}_{AB}\|_{1} \leq \Tr(\rho_{AB}) \eps \mbox{ and }
\Tr(\hat{\rho}_{AB}) \leq \Tr(\rho_{AB})\}$.
Finally, we need the related quantity of the
\emph{$\eps$-smooth min-entropy of $\rho_{AB}$ relative to $B$} defined by Renner, 
where we now
again maximize over all states $\sigma_B \in \states(\hil^B)$:
$$
\hminee{\rho_{AB}|B} \assign \sup_{\sigma_B}
\hminee{\rho_{AB}|\sigma_B}\, .
$$
We also use the shorthand
$$
\hminee{A|B} \assign \hminee{\rho_{AB}|B} \, .
$$

\subsection{Additivity}

In~\cite[Lemma 2.2]{noisy:robust} it was shown that for two independent quantum states $\rho_{X_1E_1}$ and $\rho_{X_2E_2}$
we have
$$
\hminee{X_1|E_1} + \hminee{X_2|E_2} \geq \hmin^{\eps^4}(X_1X_2|E),
$$
where $E = E_1E_2$, where $E_1$ and $E_2$ are independent. Hence, one might hope that something similar holds for a general ccq-state, in particular
that we have
\begin{equation}\label{eq:utopia}
\hminee{X_1|E} + \hminee{X_2|E} \geq \hmin^{\eps^4}(X_1X_2|E).
\end{equation}
However, we now show that there exists a cq-state 
$$
\rho_{X_1X_2E} = \sum_{x_1x_2 \in \01} p_{x_1x_2} \proj{x_1x_2} \otimes \rho_x^E,
$$
that violates this statement for small $\eps$.

From the chain-rule for the smooth min-entropy, and the data-processing 
inequality~\cite[Theorem 3.2.12]{renato:diss} we have
\begin{eqnarray*}
\hminee{X_1X_2|E} &\geq& \hminee{X_1X_2E} - H_0(E)\\
&\geq&\hminee{X_1X_2} - H_0(E)
\end{eqnarray*}
Using that $H_0(E) =  \log \rank \rho_E$ we thus have
$$
\log \dim(\hil^E) \geq \hminee{X_1X_2} - \hminee{X_1X_2|E}.
$$
Now consider the CHSH example given above. Let $p_{x_1x_2}$ be the uniform
distribution, and again let $\rho_{00} = \outp{0}{0}$, $\rho_{01} = \outp{-}{-}$, 
$\rho_{10} = \outp{+}{+}$, and $\rho_{11} = \outp{1}{1}$. The random variables $X_1$ and $X_2$
here correspond to the choice of the first and second bit respectively.

First, consider the case of $\eps = 0$. And suppose by contradiction 
that Eq.~(\ref{eq:utopia}) holds. 
Note that for our simple example we have for any $D \in \{1,2\}$ that
$\hmin(X_D|E) = - \log t$ with $t = 1/2 + 1/(2\sqrt{2})$, since the min-entropy directly relates to the
guessing probability as outlined in Eq.~(\ref{eq:duality}).
Hence, we would have
\begin{eqnarray*}
\log \dim(\hil^E) &\geq& \hmin(X_1X_2) - \hmin(X_1X_2|E)\\
&\geq&2 + 2 \log t \approx 1.54.
\end{eqnarray*}
However, we know that one qubit, i.e., $\log \dim(\hil^E) = 1$, is sufficient for this encoding.
For small $\eps$, we can make a similar argument by virtue of the fact that
$-\log [P_g(X_j|E) - \eps] \geq \hminee{X_j|E}$~\cite{rr:splitting} and $\hminee{X_1X_2} \geq \hmin(X_1X_2)$.

Additivity of the smooth min-entropy was required as a tool to show a so-called min-entropy splitting lemma for independent quantum states~\cite{noisy:robust}.
Intuitively, the technique of min-entropy splitting, first introduced by Wullschleger~\cite{Wulsc07} for classical min-entropy, states that if the min-entropy of two (or more) random variables $X_1X_2$ is high, then the min-entropy
of either $X_1$ or $X_2$ must be greater than half the joint min-entropy. Here, we are interested in the min-entropy of $X_1X_2$ \emph{conditioned} on quantum information. In particular, it was shown in~\cite[Lemma 2.7]{noisy:robust} that for $\eps \geq 0$ and two independent states $\rho_{X_1E_1}$ and $\rho_{X_2E_2}$, satisfying
$$
H^{\eps^4}_{\infty}(X_1X_2|E_1E_2) \geq \alpha,
$$
there exists a random variable $D \in \{1,2\}$ such that
$$
\hminee{X_D |E} \geq \alpha/2,
$$
with $E = E_1E_2$.
It was an open problem in~\cite{noisy:robust}, whether this statement is also true for arbitrary ccq-states
$\rho_{X_1X_2E}$.
Since additivity falls, it is no longer clear whether this would be true in general.
By the same argument as above, one can also see that for $X_1$ and $X_2$ being the random variables corresponding to the encoding
of the first or second bit respectively we cannot have that $\hmin(X_1|E) \geq \hmin(X_1X_2|E)/2$ or $\hmin(X_2|E) \geq \hmin(X_1X_2|E)/2$.

This small example shows that we must be very careful when trying to perform min-entropy splitting
with respect to quantum information, and indeed one can also use the present example to disprove min-entropy 
splitting for non-independent states. However, it does not rule out that such a statement is still true
with a significant loss in the smoothing parameter $\eps$ or by adding an additional fudge factor.
Indeed, such statements involving additional factors are known if the number of random 
variables $X_1,\ldots,X_n$ is small compared to the size of the set $\mX$ over which the 
variables $X_1,\ldots,X_n$ are distributed~\cite{rr:splitting}. Unfortunately though, they do not give
nice bounds in our setting.

\end{document}